\documentclass[12pt]{article}
\usepackage{array}
\begin{document}
\title{Bigravity in tetrad Hamiltonian formalism\\ and matter couplings}
 \author {Vladimir O. Soloviev\\
 {\small \it Institute for High Energy Physics, 142 281, Protvino, Moscow region, Russia}\\
{\small \it e-mail: Vladimir.Soloviev@ihep.ru}
}
\date{}
\maketitle
\begin{abstract}
The tetrad approach is used to resolve the matrix square root  appearing in the dRGT potential.  Constraints and their algebra are derived for the minimal case.  It is shown that the number of gravitational degrees of freedom corresponds to one massless and one massive gravitational fields when two sorts of matter separately interact with two metric tensors. The Boulware-Deser ghost is then excluded by two second class constraints. In other case when the matter couples to a linear combination of two tetrads this ghost re-appears.
\end{abstract}

\section{Introduction}
The problems of dark energy and dark matter may be addressed by modifying General Relativity (GR) at large distances. Massive gravity and bigravity~\cite{Salam,WZ,Kogan} are among the possible variants of the new theory. The potential suggested  by de Rham, Gabadadze and Toley~\cite{dRGT} (dRGT) allows to construct a model of the Universe as two extreemly weakly interacting worlds each having its own metric and a set of matter fields. When the coefficients of the dRGT-potential are going to zero each world is described by its own GR with its own sources, i.e. matter fields. So, the correspondence principle requires only small values for these coefficients and preference given to one world as known to us. It means that we know the gravitational constant and the matter content of this world. The role of the ``shadow'' world is in appearance of the massive graviton as a supplement to the massless one. Then the full number of gravitational degrees of freedom occurs
 equal to 7 (where 2 is for the massless and 5 is for the massive graviton), whereas the ghost degree of freedom~\cite{BD} is absent.  Different proofs of this statement have been developed both in the metric formalism, and in the tetrad one~\cite{HiRo,Krasnov,Alexandrov,Kluson_tetrad}.

Unfortunately, these proofs are not easy to reproduce because of their complicated form. As a rule  different notations and different variables are used by different authors.  The article~\cite{HiRo} by Hinterbichler and Rosen seems an exception that is the most readable. The authors demonstrated  advantages of the tetrad approach   in comparison to the metric one, but they did not provide a complete analysis of the constraints in bigravity. We are to stress that article~\cite{HiRo} gives a scheme (or a plan) of the proof, but not a full proof.

Some other authors later provided more detailed expositions. They were to use a lot of new variables and special tricks, and all this shadowed a bit the logical steps and the obtained results. By no means questioning the priority and correctness of previously published works we propose here a complete analysis of the Hamiltonian structure of bigravity with the dRGT-potential in tetrad variables along the scheme suggested by Hinterbichler and Rosen~\cite{HiRo}. This line of investigation is close to the approach by Kluson~\cite{Kluson_tetrad}, but differs from it by a desire to minimize the number of canonical variables and the number of constraints. Also our approach differs in the choice of variables and in some corollaries. All this will be discussed in more detail in the Conclusion.

If metric approach takes as canonical coordinates  two induced metrics $\eta_{ij}$ and $\gamma_{ij}$ with six independent components for each one,  then tetrad approach instead takes two 9-component triads, $f_{ai}$ and $e_{bj}$, so adding 6 degrees of freedom, that should be killed by  additional constraints. The other variables are not dynamical, i.e.  Lagrangian does not contain their time derivatives. This kind of variables should be determined from consistency conditions, i.e. equations providing an agreement between the Hamiltonian constraints and the dynamical equations. In the tetrad approach it is sufficient to have 3 new first class constraints and 6 new second class constraints to get the same number of independent variables as in the metric treatment. The other constraints arising in the formalism are already familar to us from the metric approach. Let us remind 
that they are one first class constraint responsible for freedom to choose a foliation of spacetime by spatial hypersurfaces, 3 first class constraints  responsible for freedom to choose spatial coordinates on these hypersurfaces, and  also  2 second class constraints   for an exclusion of the ghost degree of freedom in the dRGT potential case.  

Section 2 is devoted to introduction of notations and variables. In Section 3 the dRGT-potential is written in the proposed variables and the symmetry conditions for tetrads are given. Section 4 contains formulas for the Hamiltonian and primary constraints. In Section 5 consistency conditions are derived and resolved for auxiliary variables. Also the important secondary constraint is obtained. Section 6 is devoted to the classification of constraints and calculation of the number of gravitaional degrees of freedom. First Appendix contain explicit exposition of some short notations. In the second Appendix (Note added) we present a study of the recently proposed coupling to matter. 
  
In the present work we use the same notations as in articles~\cite{SolTch}
where bigravity Hamiltonian formalism in metric variables has been studied. An explicit form of the potential function was not used there, only  few conditions were put on it instead. Here we intend to take advantages coming from the manifest formula of dRGT-potential and compare results with the previous work~\cite{SolTch}.

At last we add a calculation of the bigravity Hamiltonian for a special form of matter-gravity coupling  proposed recently~\cite{dRHR}. It is confirmed that the constraint necessary to avoid the Boulware-Deser ghost does not appear in this case.

\section{Tetrad variables and their optimal choice}
In Hamiltonian approach we need to separate the time coordinate from the spatial ones. The state is prescribed at a spacelike hypersurface embeded in space-time. The evolution is a moving of this hypersurface through space-time, i.e. a continuous transformation of one hypersurface of state into another. So, we need a one-parametrical family of spacelike hypersurfaces. Any parameter  $t$ which numerate hypersurfaces monotonically and continuously can serve as a time. Arbitrary spatial coordinates  $x^i$ are defined at one hypersurface and continuously prolonged to the others in such a way that lines going through the points with the same coordinate values may be treated as observer worldlines, i.e. be timelike.  

In the Arnowitt-Deser-Misner approach~\cite{ADM} (ADM) the choice of a family of spacelike hypersurfaces  $t=\mathrm{const}$ and of the internal coordinates  $x^i$ is determined by the given space-time coordinates frame  $X^\mu$:
$$
t=X^0,\quad x^i=X^i,
$$
Then the metric in this coordinate basis takes a form 
\begin{equation}
 g_{\mu\nu}=\left(\begin{array}{cc}-N^2+\gamma_{mn}N^mN^n & \gamma_{jk}N^k\\\gamma_{ik}N^k & \gamma_{ij}\end{array}\right),\label{eq:ADM}
\end{equation}
where $\gamma_{ij}$ is induced metric and $N,N^i$ are  lapse and shift.

In the Kucha\u{r} approach~\cite{Kuchar} two coordinate systems are exploited, the  first one $X^\alpha$ is an arbitrary space-time frame, another $(t,x^i)$ is related to one-parametrical family of spacelike hypersurfaces, in such a way that there is exactly one hypersurface passing through any space-time point. Time  $t$ is a parameter monotonically numerating hypersurfaces, and 3 coordinates 
$x^i$ are continuously and in one-to-one way numerating points at these hypersurfaces.

Embedding functions
$$
X^{\alpha}=e^{\alpha}(x^i, t)
$$
are giving us the rules of transformation between these systems. If a choice of a hypersurfaces family
is already made, then variables 
$$
e^\alpha_i(x^k, t)\equiv\frac{\partial e^{\alpha}}{\partial x^i}
$$
are simultaneously space-time vectors and space co-vectors. The metric induced on a hypersurface of state  $t=\mathrm{const}$ is given by the following equation
$$
 \gamma_{ij}(x^k, t)=g_{\mu\nu}e^\mu_i e^\nu_j.
$$
Inverse matrices to $g_{\mu\nu}$ and $\gamma_{ij}$ are denoted correspondingly as $g^{\mu\nu}$ and $\gamma^{ij}$. They may be used for rising and lowing both Greek and Latin indices:
$$
{\bar e}_{\alpha i}=g_{\alpha\beta}e^\beta_i,\quad {\bar e}^i_{\alpha}= g_{\alpha\beta}e^\beta_j\gamma^{ij}.
$$
The bar here is introduced to distinguish variables determined with the help of metric from the initial variables $e^\alpha_i$ which are metric independent.  Below, when dealing with bigravity where another metric  $f_{\mu\nu}$ will also be involved, the bar will mark variables constructed by means of metric $g_{\mu\nu}$.

Let us introduce a co-vector normal to the hypersurface
$$
n_\alpha e^{\alpha}_i=0,
$$
and construct from it a normalized vector:
$$
{\bar n}^\alpha=g^{\alpha\beta}{\bar n}_\beta\label{eq:unitnormal}, \qquad g^{\mu\nu}{\bar n}_\mu {\bar n}_\nu=-1,
$$
with the help of metric $g_{\mu\nu}$.

Then we can decompose any space-time vectors and tensors over basis $({\bar n}^\alpha, e^\alpha_i)$.  In particular, in the Kucha\u{r} approach the lapse and shift are components of the time vector field:
\begin{equation}
\frac{\partial}{\partial t}
=Nn^\alpha+N^ie^\alpha_i. \label{eq:time}
\end{equation}

Let there are given tetrads in space-time, i.e. 4 vector fields $E^\mu_A$, $A=(0,a)$, $a=1,2,3$, orthonormalized at any point  
$$
g_{\mu\nu}E^\mu_AE^\nu_B\equiv E^\mu_AE_{\mu B}=h_{AB},
$$
where $h_{AB}=\mathrm{diag}(-1,+1,+1,+1)$, inverse matrix is denoted as $h^{AB}$. Then defining $E^A_\mu$ as $h^{AB}E_{\mu B}$  it is possible to express metric field through tetrad fields
$$
g_{\mu\nu}=E^A_\mu E^B_\nu h_{AB},\qquad g^{\mu\nu}=E^\mu_AE^\nu_Bh^{AB}.
$$  
 
In the following we will use tetrads  as the main variables both for the theory of gravity and for bigravity.

For Hamiltonian formalism the most suitable choice of a tetrad is to take unit normal $\bar n^\mu$ to  hypersurface as the timelike vector $E^\mu_0$, then other 3 vectors  $E^\mu_a$ will be tangential to the hypersurface, and so they can be related to triads that determine induced 3-metric (in analogous way as tetrads determine space-time metric if we replace $A$ by $a$ and  $h_{AB}$ by  $\delta_{ab}$)
$$
\gamma_{ij}=e^a_ie^b_j\delta_{ab}\equiv e^a_i e^a_j\equiv e_{ia}e_{ja},
$$
here $e^a_i\equiv e_{ia}$. The relation between tetrads and triads is given as follows 
$$E^\mu_a=e_{ia}\bar e^{\mu i}\equiv e_{ia}e^\mu_j\gamma^{ij},$$
and vice versa, 
$$e_{ia}=E^\mu_a\bar e_{\mu i}\equiv E^\mu_a e^\nu_ig_{\mu\nu}.$$ 
For the metric we have
\begin{equation}
g^{\mu\nu}=-E^\mu_0 E^\nu_0+E^\mu_aE^\nu_a=-\bar n^\mu \bar n^\nu+\gamma^{ij}e^\mu_i  e^\nu_j.\label{eq:g^}
\end{equation}
In covariant components
$$
E^0_\mu=-\bar n_\mu, \qquad E^a_\mu=e_i^a\bar e_\mu^i\equiv e_i^a e^\alpha_jg_{\alpha\mu}\gamma^{ij},
$$
\begin{equation}
 g_{\mu\nu}=-E^0_\mu E^0_\nu+E^a_\mu E^a_\nu=-\bar n_\mu \bar n_\nu+\bar e_\mu^i \bar e_\nu^j\gamma_{ij}.\label{eq:g_}
\end{equation}
Whereas in ADM approach space-time metric is decomposed over the space-time coordinate basis  (\ref{eq:ADM}) and have 10 components $N$, $N^i$, $\gamma_{ij}$,  in Kucha\u{r}'s formalism it is decomposed over basis $({\bar n}^\alpha, e^\alpha_i)$, where one vector is normalized and orthogonal to other vectors. Thereof the metric has only 6 nontrivial components $\gamma_{ij}$ in this decomposition, see (\ref{eq:g^}), (\ref{eq:g_}).

\section{Bigravity potential in tetrad variables} 
In metric variables the bigravity Lagrangian is a sum of two separate contributions for each metric $f_{\mu\nu}$, $g_{\mu\nu}$ having the standard General Relativity (GR) form minus the interaction potential
\begin{equation}
{\cal L}={\cal L}^{(f)}+{\cal L}^{(g)}-N\tilde U,\label{eq:Lagr}
\end{equation}
here $N\tilde U=\sqrt{-g}U$ is the density of potential which is constructed algebraically of the two metric tensors.  Here we will consider the dRGT-potential~\cite{dRGT} leading to bigravity without ghost degree of freedom~\cite{HR_bi}. Also we limit ourselves with the minimal potential case.

In the metric approach tensor  $g^{\mu\alpha}f_{\alpha\nu}$ is used to get an explicit form of the potential. It is treated as a matrix of which the matrix square root should be calculated 
$$
X^\mu_\nu=\sqrt{g^{\mu\alpha}f_{\alpha\nu}}.
$$ 
Then the dRGT-potential is defined as a linear combination 
$$
U=\sum_{i=0}^{4}\beta_i U_i
$$
of symmetric polynomials formed from eigenvalues of matrix  $X^\mu_\nu$, which could be expressed throgh traces of different powers of this matrix:
\begin{eqnarray}
U_0&=& 1,\nonumber\\
 U_1&=& \mathrm{Tr}X,\nonumber\\
 U_2&=& \frac{1}{2}\left((\mathrm{Tr}X)^2-\mathrm{Tr}X^2\right),\nonumber\\
 U_3&=& \frac{1}{6}\left((\mathrm{Tr}X)^3-3\mathrm{Tr}X\mathrm{Tr}X^2+2\mathrm{Tr}X^3\right),\nonumber\\
U_4&=&\frac{\sqrt{-f}}{\sqrt{-g}}.\nonumber
\end{eqnarray}
In the present work we consider the minimal dRGT potential only:
\begin{equation}
\sqrt{-g}U=\beta_0\sqrt{-g}+\beta_1\sqrt{-g}U_1+\beta_4\sqrt{-f}.\label{eq:minpot}
\end{equation} 
It is evident that only $U_1$ is responsible for the interaction of the two metrics, other contributions simply modify the two cosmological terms $\Lambda^{(f)}$, $\Lambda^{(g)}$. 

The main difficulty in constructing the Hamiltonian formalism in metric variables is a problem how to find analytically the matrix square root. In publication~\cite{HR_bi} a complicated matrix transformation was proposed to deal with this problem. With the help of it the authors of~\cite{HR_bi} found two additional second class constraints sufficient to exclude the ghost degree of freedom. In some other works Stuckelberg fields were used  for this purpose. One more direction of attack~\cite{SolTch,Comelli_MayDay} was to start with the potential of a general form and then find the conditions necessary and sufficient to exclude the ghost.  But the most straightforward direction was proposed by Hinterbichler and Rosen~\cite{HiRo} where it was suggested to use tetrad variables for explicit calculation of the square root.

In bigravity we need to double the set of terad variables 
$$
g^{\mu\nu}=E^\mu_AE^\nu_Bh_{AB},\qquad  f^{\mu\nu}=F^\mu_AF^\nu_Bh_{AB},
$$
which should be acompanied by the symmetricity conditions 
\begin{equation}
E^\mu_AF_\mu^B-E^{\mu B}F_{\mu A}=0.\label{eq:condition}
\end{equation}
In fact, under these conditions a solution of equation 
$$
 g^{\mu\alpha}f_{\alpha\nu}=X^\mu_\beta X^\beta_\nu,
$$
is the following matrix
$$
X^\mu_\nu=E^{\mu A}F_{\nu A}.
$$

Unfortunately, we cannot take as a second tetrad  variables $F_{\nu A}$, constructed in a similar way to to $E^\mu_A$  (in previous Section), i.e. as components of the optimal tetrad manufactured now with metric  $f_{\mu\nu}$. If potential $U$ is expressed through two metric tensors then it is invariant under rotations of each tetrad separately. But if it is expressed through two  tetrads with the symmetricity conditions satisfied, then it is invariant only under diagonal space-time rotations.  
$$
E^{\mu A}\rightarrow \Lambda^A_{\ B} E^{\mu B}, \qquad F_{\nu B}\rightarrow F_{\nu B}\Lambda^B_{\ A}.
$$
Therefore all we can do is to take one tetrad as optimal and another as general. As it was mentioned in article~\cite{HiRo}, any general tetrad can be obtained as a boost transformation of some optimal one
  \begin{equation}
\Lambda^A_{ \ B}=\left(\begin{array}{cc} \varepsilon & p_b \\
p^a & \delta^a_{ \ b}+\frac{1}{\varepsilon+1}p^a p_b \\ \end{array}
\right), \qquad \varepsilon=\sqrt{1+p_ap^a}\label{eq:Lorentz} \ ,
\end{equation}
where $$
p^a=p_a,\qquad
h_{AB}\Lambda^A_{ \ C}\Lambda^B_{ \ D}=h_{CD}.
$$
Then a general form of matrix $X^\mu_\nu$ is the following
$$
X^\mu_\nu=E^\mu_A F^A_\nu, \qquad\mbox{where}\qquad F^A_\nu=\Lambda^A_{\ B}(p) {\cal F}^B_\nu,\label{eq:matrixX}
$$
wherein
$$
E^\mu_0=\bar n^\mu,\qquad {\cal F}_\mu^0=-n_\mu,
$$
$$
E^\mu_a=e_{ia}\gamma^{ij}e_j^\mu,\qquad {\cal F}_\mu^a=f_{ia}\eta^{ij}e^\beta_j f_{\beta\mu},
$$
\begin{equation}
\gamma_{ij}=e^a_ie^b_j\delta_{ab},\qquad \eta_{ij}=f^a_if^b_j\delta_{ab},\label{eq:ef}
\end{equation}
and the trace of matrix $X^\mu_\nu$ is defined as follows
\begin{eqnarray}
\mathrm{Tr}X&=&E^\mu_A\Lambda^A_{\ B}(p) {\cal F}^B_\mu\nonumber\\
&=&-\varepsilon n_\mu\bar n^\mu +f_{\mu\nu}\bar n^\mu e^\nu_j(p_af^{ja})+\gamma^{ij}(e_{ia}f_{j}^a)+\frac{\gamma^{ij}(e_{ia}p^a)(f_j^bp_b)}{\varepsilon+1}.\nonumber
\end{eqnarray}
Given a relation between the two bases derived in articles~\cite{SolTch}
\begin{equation}
 \bar n^\mu=\sqrt{-g^{\perp\perp}}n^\mu-\frac{g^{\perp i}}{\sqrt{-g^{\perp\perp}}}e^\mu_i,\label{eq:bases}
\end{equation}
and applying the definition of variables $u,u^i$ given there 
$$
u=\frac{1}{\sqrt{-g^{\perp\perp}}},\qquad u^i=-\frac{g^{\perp i}}{g^{\perp\perp}},
$$
where
$$
g^{\perp\perp}=g_{\mu\nu}n^\mu n^\nu, \qquad g^{\perp i}=-g_{\mu\nu}n^\mu e^{\nu i},
$$
we obtain the following expression for the minimal potential (\ref{eq:minpot})
$$
\tilde U=\beta_0 u\sqrt{\gamma}+\beta_1\sqrt{\gamma}\left(\varepsilon +u(f^a_ie^i_a)+\frac{u}{\varepsilon+1}(f^a_i p_a)(e^i_b p^b)- u^i (f^a_ip_a)\right)+\beta_4\sqrt{\eta},
$$
or
\begin{equation}
\tilde U=\beta_0 ue+\beta_1e\left[\varepsilon+uz - (ufp)\right]+\beta_4 f,\label{eq:Pot}
\end{equation}
where we have introduced the following notations:
$$
x_{ab}=f^a_ie^i_b,\quad y_{ab}=p_ap_cx_{cb},  \quad z_{ab}=x_{ab}+\frac{y_{ab}}{\varepsilon+1},\quad (ufp)=u^i f^a_ip_a,
$$
$$
e=\det ||e_{ia}||,\quad f=\det ||f_{ia}||, \quad x=x_{aa},\quad y =y_{aa},\quad z=z_{aa}.
$$
It is useful also to introduce a notation $x^{ij}=f^{ia}e^{ja}$. Variables $x_{ab}$, $x^{ij}$ may serve as transfer matrices between triads:
\begin{eqnarray}
 f_a^ix^{ab}&=&e^{ib},\qquad x_{ab}e_i^b=f_{ia},\nonumber\\
f_i^ax^{ij}&=&e^{ja},\qquad x^{ij}e_j^a=f^{ia}.\nonumber
\end{eqnarray}
Potential (\ref{eq:Pot}) is linear in variables $u,u^i$, and so evidently satisfies the homogeneous Monge-Amp\`ere equation, 
$$
\det \left|\left| \frac{\partial^2\tilde U}{\partial u^a\partial u^b}\right|\right|=0,
$$
that is a necessary condition for exclusion of ghosts as discussed in~\cite{SolTch,Comelli_MayDay} . The Hessian rank which has been equal to 3 in metric approach~\cite{SolTch} here is equal to zero.

The symmetry conditions 
(\ref{eq:condition})
give us 6 equations which can be rewritten as follows
\begin{eqnarray}
G_a&\equiv&p_a+up_bx_{ba}-u^jf_j^b\left(\delta_{ab}+\frac{p_ap_b}{\varepsilon+1}\right)=0,\label{eq:const1}\\
G_{ab}&\equiv&x_{ab}-x_{ba}+\frac{p_c}{\varepsilon+1}\left(p_ax_{cb}-p_bx_{ca} \right)\equiv z_{[ab]}=0,\label{eq:const2}
\end{eqnarray}
where square brackets denote antisymmetrization of indices, and round brackets below will be used for symmetrization.  The first condition may be interpreted as fixing variable
 $u^i$
\begin{equation}
 u^i=\frac{p_af^{ib}}{\varepsilon}\left[\delta_{ab}+u\varepsilon\left( x_{ab} -\delta_{ab}\frac{y}{\varepsilon(\varepsilon+1)}\right)\right].\label{eq:ui_0}
\end{equation}
The second condition finally will occur a constraint on canonical variables.

\section{Hamiltonian and primary constraints}
In metric approach the bigravity Lagrangian (\ref{eq:Lagr}) is a sum of two separate contributions for each metric minus the interaction potential, each contribution is a Lagrangian of GR. 
In a similar way, the Hamiltonian of bigravity (given the tetrad symmetricity conditions (\ref{eq:const1}), (\ref{eq:const2})) 
can be written as follows
\begin{equation}
{\rm H}={\rm H}^{(f)}+{\rm H}^{(g)}+\int d^3x \left(N\tilde U+\Lambda^aG_a+\Lambda^{ab}G_{ab}\right).\label{eq:H}
\end{equation}
 We take 36 functions $(f_{ia},\Pi^{ia})$, $(e_{ia},\pi^{ia})$ as bigravity canonical variables in tetrad approach, wherein each Hamiltonian  ${\rm H}^{(f)}$, ${\rm H}^{(g)}$ contains primary constraints providing the freedom to choose triads $e^a_i$ и $f^a_i$, 
\begin{eqnarray}
 L_{ab}&=&f_{ia}\Pi^i_b-f_{ib}\Pi^i_a=0,\label{eq:Labf}\\
 \bar L_{ab}&=&e_{ia}\pi^i_b-e_{ib}\pi^i_a=0.\label{eq:Labg}
\end{eqnarray}
Every constraint appears in the action with its own Lagrangian multiplier:
\begin{eqnarray}
{\rm H}&=&
\int d^3x \left(N{\cal H}+N^i{\cal H}_i+\lambda^{ab}L_{ab}\right)+\nonumber\\
&+&
\int d^3x \left(\bar N\bar{\cal H}+\bar N^i\bar{\cal H}_i+\bar\lambda^{ab}\bar L_{ab}\right)+\nonumber\\
&+&\int d^3x \left(N\tilde U+\Lambda^aG_a+\Lambda^{ab}G_{ab}\right).\nonumber
\end{eqnarray}
The following formulas complement relations (\ref{eq:ef}) between canonical variables of the metric and tetrad formalisms:
\begin{eqnarray}
 \Pi^{ij}&=&\frac{1}{4}\left(f^{ia}\Pi^{ja}+f^{ja}\Pi^{ia} \right),\nonumber\\
 \pi^{ij}&=&\frac{1}{4}\left(e^{ia}\pi^{ja}+e^{ja}\pi^{ia} \right).\nonumber
\end{eqnarray}
Poisson brackets for the tetrad variables have the following form
\begin{eqnarray}
\{f_{ia}(x),\Pi^{jb}(y)\}&=&\delta_a^{b}\delta_i^{j}\delta(x,y)=\{e_{ia}(x),\pi^{jb}(y)\},\nonumber\\
\{f_{ia}(x),e_{jb}(y)\}&=&0=\{\pi^{ia}(x),\Pi^{jb}(y)\}=\{f_{ia}(x),f_{jb}(y)\},\nonumber\\
\{e_{ia}(x),e_{jb}(y)\}&=&0=\{\pi^{ia}(x),\pi^{jb}(y)\}=\{\Pi^{ia}(x),\Pi^{jb}(y)\}.\nonumber
\end{eqnarray}
Surely, we may call variables $f_{ia},\Pi^{ia}$, $e_{ia},\pi^{ia}$ not tetrad, but triad ones. As discussed above, the supplement of a unit normal vector of the hypersurface transforms a triad into an optimal tetrad, and Lorentz transformations  (\ref{eq:Lorentz})  makes of it a general tetrad.

Poisson brackets between the metric formalism momenta expressed through triad (or tetrad) variables are equal to zero on the constraint surface only, i.e. in weak sense
\begin{eqnarray}
 \{\Pi^{ij}(x),\Pi^{k\ell}(y)\}&=&\frac{1}{4}\left(\eta^{ik}{\cal M}^{j\ell}+\eta^{i\ell}{\cal M}^{jk}+\eta^{jk}
{\cal M}^{i\ell}+\eta^{j\ell}{\cal M}^{ik} \right)\approx 0,\nonumber\\
\{\pi^{ij}(x),\pi^{k\ell}(y)\}&=& \frac{1}{4}\left(\gamma^{ik}\bar{\cal M}^{j\ell}+\gamma^{i\ell}\bar{\cal M}^{jk}+\gamma^{jk}\bar{\cal M}^{i\ell}+\gamma^{j\ell}\bar{\cal M}^{ik} \right)\approx 0,\nonumber
\end{eqnarray}
where
\begin{eqnarray}
 {\cal M}^{ij}&=&\frac{1}{4}L_{ab}f^{ja}f^{ib}\equiv\frac{1}{4}\left(f^{ia}\Pi^{j}_a-f^{ja}\Pi^i_a \right)\approx 0, \nonumber\\
\bar{\cal M}^{ij}&=&\frac{1}{4}\bar L_{ab}e^{ja}e^{ib}\equiv\frac{1}{4}\left(e^{ia}\pi^{j}_a-e^{ja}\pi^i_a \right)\approx 0.\nonumber
\end{eqnarray}
Similarly, Poisson brackets of constraints ${\cal H},{\cal H}_i$ and $\bar{\cal H},\bar{\cal H}_i$ in the tetrad variables differ from those given below (\ref{eq:bialg1}) -- (\ref{eq:bialg3}), (\ref{eq:bialg4}) -- (\ref{eq:bialg6}),  by contributions proportional to ${\cal M}^{ik}$ and $\bar{\cal M}^{ik}$ respectively.
In two copies of GR with metrics  $f_{\mu\nu}$  and  $g_{\mu\nu}$ respectively the Poisson brackets of expressions  ${\cal H}$, ${\cal H}_i$ and $\bar{\cal H}$, $\bar{\cal H}_i$, which are in this case constraints,  have the following form
 \begin{equation}
 \{ {\cal H}(x),{\cal H}(y)\}=\left[\eta^{ik}(x){\cal H}_k(x)+ \eta^{ik}(y){\cal H}_k(y)\right]\delta_{,i}(x,y),\label{eq:bialg1}
\end{equation}
\begin{equation}
  \{ {\cal H}_i(x),{\cal H}_k(y)\}={\cal H}_i(y)\delta_{,k}(x,y)+ {\cal H}_k(x)\delta_{,i}(x,y),\label{eq:bialg2}
\end{equation}
\begin{equation}
  \{ {\cal H}_i(x),{\cal H}(y)\}={\cal H}(x)\delta_{,i}(x,y).\label{eq:bialg3}
\end{equation}
 In bigravity these brackets are given by the same formulas, but now the expressions are not constraints, and apart from (\ref{eq:bialg1}), (\ref{eq:bialg2}), (\ref{eq:bialg3}), there are similar relations 
\begin{equation}
 \{ \bar{\cal H}(x),\bar{\cal H}(y)\}=\left[\gamma^{ik}(x)\bar{\cal H}_k(x)+ \gamma^{ik}(y)\bar{\cal H}_k(y)\right]\delta_{,i}(x,y),\label{eq:bialg4}
\end{equation}
\begin{equation}
  \{ \bar{\cal H}_i(x),\bar{\cal H}_k(y)\}=\bar{\cal H}_i(y)\delta_{,k}(x,y)+ \bar{\cal H}_k(x)\delta_{,i}(x,y),\label{eq:bialg5}
\end{equation}
\begin{equation}
  \{ \bar{\cal H}_i(x),\bar{\cal H}(y)\}=\bar{\cal H}(x)\delta_{,i}(x,y).\label{eq:bialg6}
\end{equation}
for $\bar{\cal H},\bar{\cal H}_i$. In tetrad variables the above relations are valid modulo constraints $L_{ab}$ and $\bar L_{ab}$, respectively. 
In the metric approach the following formulas take place
\begin{eqnarray}
{\cal  H}&=&{\cal H}_M-\frac{\sqrt{\eta}}{\kappa^{(f)}} \left(R^{(\eta)}-2\Lambda^{(f)}\right)-\frac{\kappa^{(f)}}{\sqrt{\eta}}\left(\frac{\Pi^2}{2}-\mathrm{Tr}\Pi^2  \right),\nonumber\\
{\bar{\cal  H}}&=&{\bar{\cal H}}_M-\frac{\sqrt{\gamma}}{\kappa^{(g)}}\gamma \left(R^{(\gamma)}-2\Lambda^{(g)}\right)-\frac{\kappa^{(g)}}{\sqrt{\gamma}}\left(\frac{\pi^2}{2}-\mathrm{Tr}\pi^2 \right),\nonumber 
\end{eqnarray}
    and 
 \begin{eqnarray}   
 {\cal  H}_i&=&{\cal H}_{iM}-2\Pi_{i|j}^j,\nonumber\\
  {\bar  {\cal  H}}_i&=&{\bar{\cal H}}_{iM}-2\pi_{i|j}^j,\nonumber
 \end{eqnarray}   
where ${{\cal H}}_M, {{\cal H}}_{iM}$, ${\bar{\cal H}}_M, {\bar{\cal H}}_{iM}$ are contributions of the two types of matter.

As variables $N,N^i$, $\bar N,\bar N^i$ are components of the same time-vector field   (\ref{eq:time}) decomposed over two different bases $(n^\alpha,e^\alpha_i)$ $(\bar n^\alpha,e^\alpha_i)$, the relation between the bases (\ref{eq:bases}) determines the relation between components:
$$
 \bar N=Nu,\qquad \bar N^i=N^i+Nu^i.
$$
Besides $(f_{ia},\Pi^{ia})$, $(e_{ia},\pi^{ia})$ there are noncanonical variables in the Hamiltonian, their characteristic feature is that their velocities do not appear in the Lagrangian.
Let us mention that variables
 $N$, $N^i$, $u$, $u^i$, $\lambda^{ab}$, $\bar\lambda^{ab}$, $\Lambda^a$, $\Lambda^{ab}$ appears in the Hamiltonian only in linear way.
The variation of action in these variables leads to new equations. ~\footnote{In article~\cite{Kluson_tetrad} the action is varied also over variable $p_a$, that gives 3 other  equations to determine $u^i$ (in work~\cite{Kluson_tetrad} the corresponding variable is denoted as $n^i$) instead of our equations (\ref{eq:const1}). The latter are a half of symmetry conditions  (\ref{eq:condition}) which are necessary for equivalence of the tetrad and the metric formulations of bigravity. The second half of symmetry conditions (\ref{eq:const2}) is present in work~\cite{Kluson_tetrad} where they arise as  secondary constraints.} We will see below that only part of these equations are primary constraints on canonical variables, the rest serves to find some of these auxiliary variables. It is suitable for the following narration to replace constraints
 (\ref{eq:Labf}), (\ref{eq:Labg}) by their symmetric and antisymmetric combinations:
\begin{eqnarray}
L^+_{ab}&\equiv&L_{ab}+\bar L_{ab}=0,\nonumber\\
L^-_{ab}&\equiv&L_{ab}-\bar L_{ab}=0,\nonumber
\end{eqnarray}
replacing respectively also the Lagrangian multipliers: 
 \begin{eqnarray}
\lambda^+_{ab}&=&\frac{1}{2}\left(\lambda_{ab}+\bar\lambda_{ab}\right),\nonumber\\
\lambda^-_{ab}&=&\frac{1}{2}\left(\lambda_{ab}-\bar \lambda_{ab}\right).\nonumber
\end{eqnarray}
By varying the Hamiltonian 
\begin{eqnarray}
{\rm H}&=&\int d^3x \left(
N({\cal H}+u\bar{\cal H}+u^i\bar{\cal H}_i+\tilde U)+N^i({\cal H}_i+\bar{\cal H}_i)\right.+\nonumber\\
&+&\left.\lambda^+_{ab}L^+_{ab}+ \lambda^-_{ab}\bar L^-_{ab}+\Lambda^aG_a+\Lambda^{ab}G_{ab}\right).\label{eq:H_0}
\end{eqnarray}
over $\lambda^+_{ab}$, $\lambda^-_{ab}$ we obtain the following constraints 
\begin{eqnarray}
L^+_{ab}&\equiv&f_{ia}\Pi^i_b-f_{ib}\Pi^i_a+e_{ia}\pi^i_b-e_{ib}\pi^i_a=0,\nonumber\\
L^-_{ab}&\equiv&f_{ia}\Pi^i_b-f_{ib}\Pi^i_a-e_{ia}\pi^i_b+e_{ib}\pi^i_a=0,\nonumber
\end{eqnarray}
by varying over $\Lambda^a$, $\Lambda^{ab}$ we get symmetry conditions (\ref{eq:const1}), (\ref{eq:const2}),
varying over $u$ gives us the following (given  (\ref{eq:const1}), (\ref{eq:const2}))
\begin{equation}
{\cal S}\equiv\frac{1}{N}\frac{\delta{\rm H}}{\delta u}= \bar{\cal H}+\frac{\partial\tilde U}{\partial u}=\bar{\cal H}+\beta_1 e z
=0,\label{eq:S}
\end{equation}
and by varying over $u^i$, given (\ref{eq:const1}), (\ref{eq:const2}), we obtain
\begin{equation}
 {\cal S}_i\equiv \frac{1}{N}\frac{\delta{\rm H}}{\delta u^i}=\bar{\cal H}_i+\frac{\partial\tilde U}{\partial u^i}=\bar{\cal H}_i-\beta_1ef^a_ip_a=0,\label{eq:Si}
\end{equation}
It is possible to find  $p_a$ from the last equation:
\begin{equation}
p_a=\frac{1}{\beta_1e}f_a^i\bar {\cal H}_i,\label{eq:pa}
\end{equation}
then
$$
\varepsilon\equiv\sqrt{1+p_ap^a}=\frac{\sqrt{(\beta_1 e)^2+\eta^{mn}\bar{\cal H}_m\bar{\cal H}_n}}{\beta_1 e},
$$
and therefore,  it follows from (\ref{eq:ui_0})
\begin{equation}
u^i= \frac{\bar{\cal H}_k}{\beta_1 e\varepsilon}\left[\eta^{ik}+u\varepsilon \left(x^{ik}-\eta^{ik}\frac{y}{\varepsilon(\varepsilon+1)}\right)\right].\label{eq:ui}
\end{equation}
Varying bigravity action over $N$, $N^i$ we get 
\begin{eqnarray}
{\cal R}&\equiv&{\cal H}+u\bar{\cal H}+u^i\bar{\cal H}_i+\tilde U={\cal H}+u{\cal S}+u^i{\cal S}_i+\beta_1 e\varepsilon=0,\label{eq:R}\\
{\cal R}_i&\equiv&{\cal H}_i+\bar{\cal H}_i=0,\label{eq:Ri}
\end{eqnarray}
As we are able to exclude auxiliary variables $p_a$, $u^i$ by solving equations (\ref{eq:Si}) and  (\ref{eq:const1}),  the Hamiltonian takes a form
$$
 {\rm H}=\int d^3x \left(N({\cal R}'+u{\cal S})+N^i{\cal R}_i+\lambda^+_{ab}L^+_{ab}+\lambda^-_{ab}L^-_{ab}+\Lambda_{ab}G_{ab}\right).
$$
where the full set of primary constraints are as follows
\begin{eqnarray}
{\cal R}'&\equiv&{\cal H}+\beta_1 e \varepsilon=0,\\
{\cal R}_i&\equiv&{\cal H}_i+\bar{\cal H}_i=0,\\
{\cal S}&=&\bar{\cal H}+\beta_1ez=0,\label{eq:S2}\\
G_{ab}&\equiv&z_{[ab]}=0,\label{eq:Gab}\\
L^+_{ab}&\equiv&f_{ia}\Pi^i_b-f_{ib}\Pi^i_a+e_{ia}\pi^i_b-e_{ib}\pi^i_a=0,\label{eq:L+}\\
L^-_{ab}&\equiv&f_{ia}\Pi^i_b-f_{ib}\Pi^i_a-e_{ia}\pi^i_b+e_{ib}\pi^i_a=0.\label{eq:L-}
\end{eqnarray}

\section{Consistency of primary constraints and dynamics}
We should provide the preservation of primary constraints in the course of evolution, and for this purpose it is necessary to calculate the Poisson brackets between the constraints.
If in short we denote as $L^{\cal A}$  any constraint from the set $L_{ab},\bar L_{ab},L^+_{ab},L^-_{ab}$ and as  $H_{\cal B}$ any expression like ${\cal H}$, $\bar{\cal H}$, ${\cal H}_i$, $\bar{\cal H}_i$, $e$, $f$, then all Poisson brackets $\{L^{\cal A}, H_{\cal B}\}$ will be weakly zero, because they are linear combinations of constraints $L^{\cal A}$. 
As a result we obtain
\begin{eqnarray}
 \{L^+_{ab},{\cal S}\}&\approx&\beta_1e\{L^+_{ab},z\}\approx 0,\nonumber\\
\{L^-_{ab},{\cal S}\}&\approx&\beta_1e\{L^-_{ab},z\}\approx -2\beta_1 eG_{ab}\delta(x,y)\approx 0,\nonumber\\
 \{L^+_{ab},G_{cd}\}&=&\left[\delta_{ac}G_{bd}+\delta_{ad}G_{cb}-\delta_{bc}G_{ad}-\delta_{bd}G_{ca}\right]\delta(x,y)\approx 0,\nonumber\\
\{L^-_{ab},G_{cd}\}&=&\left[\delta_{ac}z_{(bd)}-\delta_{ad}z_{(cb)}-\delta_{bc}z_{(ad)}+\delta_{bd}z_{(ca)}\right]\delta(x,y)\ne 0,\nonumber\\
\{{\cal R}'(x),{\cal R}'(y)\}&=&\left[\eta^{ik}(x){\cal R}_i(x)+\eta^{ik}(y){\cal R}_i(y)\right]\delta_{,k}(x,y)\approx 0,\nonumber\\
\{{\cal R}'(x),{\cal R}_i(y)\}&=&-{\cal R}'(y)\delta_{,i}(y,x)\approx 0,\nonumber\\
\{{\cal S}(x),{\cal R}_i(y)\}&=&-{\cal S}(y)\delta_{,i}(y,x)\approx 0,\nonumber\\
\{{\cal S}(x),{\cal S}(y)\}&=& J^i(x){\cal S}(x)\delta_{,i}(x,y)- J^i(y){\cal S}(y)\delta_{,i}(y,x)\approx 0,\nonumber\\
\{{\cal S}(x),{\cal R}'(y)\}&=&- K^i(y){\cal S}(y)\delta_{,i}(y,x)+\Omega\delta(x,y)\approx\Omega\delta(x,y),\label{eq:algebra}
\end{eqnarray}
where expressions $J^i$, $K^i$ are given by formulas (\ref{eq:JK}).
In addition there appears a new expression
\begin{equation}
\Omega=\frac{\partial{\cal R}'}{\partial\Pi^{ia}}\frac{\partial{\cal S}}{\partial f_{ia}}-\frac{\partial{\cal S}}{\partial\pi^{ia}}\frac{\partial{\cal R}'}{\partial e_{ia}}+\Omega',
\label{eq:Omega}
\end{equation}
where $\Omega'$ is given in equation (\ref{eq:Omega'}).
Given equations (\ref{eq:algebra}) we obtain that on the surface of the primary constraints the following relations are valid
\begin{eqnarray}
\dot {\cal R}'&=&\{ {\cal R}', {\rm H}\}\approx -Nu\Omega ,\\
\dot {\cal R}_i&=& \{{\cal R}_i, {\rm H}\}\approx 0 ,\\
\dot {\cal S}&=&\{{\cal S} , {\rm H}\}\approx Nu\Omega ,\\
\dot G_{ab}&=& \{G_{ab}, {\rm H}\}\approx \lambda^-_{cd}\left[G_{ab},L^-_{cd}\right]+\int d^3x \Lambda_{cd}(x)\{G_{ab},G_{cd}(x)\},\label{eq:evolGab}\\
\dot L^+_{ab}&=& \{L^+_{ab}, {\rm H}\}\approx 0 ,\\
\dot L^-_{ab}&=& \{L^-_{ab},{\rm H}\}\approx \Lambda_{cd}\left[L^-_{ab},G_{cd}\right],\label{eq:evolL-}
\end{eqnarray}
where square brackets denote coefficients standing before $\delta$-function in the corresponding ultralocal Poisson brackets, for example,
$$
 \{L^-_{ab},G_{cd}\}  \equiv\left[L^-_{ab},G_{cd}\right]\delta(x,y).
$$
Given nondegeneracy of matrix $||\{L^-_{ab},G_{cd}\}||$, evident from equations (\ref{eq:algebra}), we conclude that constraints $L^-_{ab}$, $G_{cd}$ are second class. To preserve the primary constraints in the course of evolution the last equation demands  $\Lambda_{cd}=0$.  Then from equation (\ref{eq:evolGab}) it follows $\lambda^-_{cd}=0$, and from the first and third equations we obtain a secondary constraint
$$
 \Omega=0.
$$
After substitution of the Lagrangian multipliers found above, bigravity Hamiltonian is as follows
$$
 {\rm H}=\int d^3x \left(N({\cal R}'+u{\cal S})+N^i{\cal R}_i+\lambda^+_{ab}L^+_{ab}\right). 
$$
The condition of consistency  of secondary constraint $\Omega$ and dynamics leads to equation for auxiliary variable $u$:
\begin{equation}
 \dot\Omega=\{\Omega,{\rm H}\}=\int d^3x N\{\Omega,{\cal R}\}=\int d^3x \left( N\{\Omega,{\cal R}'\}+ Nu\{\Omega,{\cal S}\}
\right)=0.\label{eq:find_u}
\end{equation}
The following two points are important: first, Poisson bracket $\{\Omega,{\cal S}\}$ should be nonzero, second, if we wish that variable  $u$ does not depend on Lagrangian multiplier $N$ (this is not a necessary requirement), then this bracket together with Poisson bracket $\{\Omega,{\cal R}'\}$ should be ultralocal, i.e. having a following form
\begin{equation}
\{\Omega,{\cal S}\}=\left[\Omega,{\cal S}\right]\delta(x,y),\qquad \{\Omega,{\cal R}'\}=\left[\Omega,{\cal R}'\right]\delta(x,y).\label{eq:OmS}
\end{equation}
We may see a supporting argument to the first point in the fact that in calculating  $\{\Omega,{\cal S}\}$  we obtain algebraic expressions bilinear in momenta conjugate to different metrics:
$$
\beta_1\frac{\partial^2ez}{\partial f_{ia}\partial e_{jb}}\frac{\partial{\cal H}}{\partial\Pi^{ia}}\frac{\partial\bar{\cal H}}{\partial\pi^{jb}}.
$$
There are no such expressions in other constraints. 
Second, as constraint $\Omega$ arises in the course of calculation of Poisson bracket $\{{\cal S},{\cal R}'\}$, we can apply the Jacobi identity to derive the following relations
\begin{eqnarray}
\{\Omega(x),{\cal S}(z)\}\delta(x,y)-\{\Omega(z),{\cal S}(x)\}\delta(z,y)&\approx& 0,\nonumber\\
\{\Omega(y),{\cal R}'(z)\}\delta(x,y)-\{\Omega(z),{\cal R}'(y)\}\delta(x,z)&\approx& 0.\nonumber
\end{eqnarray}
 Therefore the odd derivatives of $\delta$-function are absent in the mentioned brackets, and if we were able to prove the absence of $\delta$-function second derivatives, then our second point, i.e. conditions (\ref{eq:OmS})   would be fulfilled.  In that case we would have a simple result
$$
 u=-\frac{\{\Omega,{\cal R}'\}}{\{\Omega,{\cal S}\}}.\label{eq:ufinal}
$$
Unfortunately up to now we are unable to prove this.

\section{Classification of constraints}
Eventually this is a list of the first class constraints:
\begin{eqnarray}
{\cal R}'&\equiv&{\cal H}+\beta_1 e\varepsilon=0,\nonumber\\
{\cal R}_i&\equiv&{\cal H}_i+\bar{\cal H}_i=0,\nonumber\\
L^+_{ab}&\equiv&f_{ia}\Pi^i_b-f_{ib}\Pi^i_a+e_{ia}\pi^i_b-e_{ib}\pi^i_a=0,\nonumber
\end{eqnarray}
the corresponding Lagrangian multipliers $N$, $N^i$, $\lambda^+_{ab}$ are arbitrary according to the invariance of the formalism under diagonal diffeomorphisms and diagonal rotations of the spatial triads.

Constraints $L^-_{ab}$, $G_{cd}$, ${\cal S}$, $\Omega$ are second class:
$$
\{L^-_{ab},G_{cd}\}\ne 0,\qquad
\{\Omega,{\cal S}\}\ne 0,
$$
wherein ${\cal S}$ and $\Omega$ are responsible for exclusion of the ghost degree of freedom, analogously to the corresponding constraints in the metric approach~\cite{SolTch}. 

The constraints may be also characterized in their dependency of gravitational momenta $\Pi^{ia}$, $\pi^{ia}$.
Derivatives of momenta are present only in expressions ${\cal H}_i$, $\bar{\cal H}_i$, which in their turn appear linearly in first class constraints ${\cal R}_i$ and in expression $p_a$. Algebraic dependence on momenta in constraints ${\cal R}'$, ${\cal S}$ is quadratic, and in constraints $L^+_{ab}$, $L^-_{ab}$, $\Omega$ it is linear.  

If we will be able to prove that Poisson brackets of constraints  (\ref{eq:OmS}) are ultralocal, then it will be easy to construct the corresponding Dirac brackets, and exclude all the second class constraints from the Hamiltonian. In that case the dynamical equations will be generated by the Hamiltonian containing only first class constraints multiplied by the corresponding arbitrary Lagrangian multipliers. 
The bigravity Hamiltonian in tetrad approach for the minimal dRGT potential in that case will be as follows
$$
 {\rm H}=\int d^3x \left(N{\cal R}'+N^i{\cal R}_i+\lambda^+_{ab}L^+_{ab}\right). 
$$
i.e. it will be a linear combination of $n_{f.c.}=7$ first class constraints with arbitrary Lagrangian multipliers.

Without reference to these assumptions,  canonical variables of the constructed formalism are the two sets of triads  $f_{ia}$, $e_{ia}$ and their conjugate momenta    $\Pi^{ia}$, $\pi^{ia}$,
i.e. $n=36$ variables. Besides the first class constraints, these variables are complied to $n_{s.c.}=8$ second class constraints
$$
 {\cal S}=0,\quad G_{ab}=0,\quad
L^-_{ab}=0,\quad\Omega=0,
$$
their form is given by relations (\ref{eq:S2}), (\ref{eq:Gab}), (\ref{eq:L-}) and (\ref{eq:Omega}) respectively.
The number of gravitational degrees of freedom is calculated by the following formula
$$
n_{\mathrm{DOF}}=\frac{1}{2}\left(n-2n_{f.c.}-n_{s.c.} \right)=7. 
$$
Other variables appeared in the initial Hamiltonian (\ref{eq:H_0}), as it has been shown above, are uniquely determined through the canonical variables and possibly also through Lagrangian multiplier  $N$: 
 $p_a$  is given by equation (\ref{eq:pa}),  $u^i$ is found from equation (\ref{eq:ui}), and $u$  from (\ref{eq:find_u}). Multipliers $\lambda^-_{ab}$, $\Lambda_{ab}$, as it has been shown above, are zero as follows from (\ref{eq:evolGab}), (\ref{eq:evolL-}). 
 \begin{figure}
\begin{tabular}{|c||c|c|c|c|c|c|c|}
\hline 
variable  & equation
 & $\rightarrow$ 
 & 
result 1 & $\rightarrow$ 
 & 
 result 2& $\rightarrow$ 
 & result 3
 \\ 
\hline \hline
$N$ & ${\cal R}\approx 0$ &  &  &   &  &  & \\
\hline
$N^i$ & ${\cal R}_i\approx 0$  &  & &  & &  &  \\
\hline
$\lambda^+_{ab}$  & $L^+_{ab}\approx 0$ & & &  &  &  &   \\
\hline
$u^i$ & ${\cal S}_i=0$  &$\rightarrow$  & $p_a$ &  &  &  &  \\
\hline
$\Lambda^a$ & $G_a=0$ & $\rightarrow$ & $u^i$ &   &  &  &   
\\
\hline
$\lambda^-_{ab}$ & $L^-_{ab}=0$ & $\rightarrow$ & $\{L^-_{ab},G_{cd}\}\ne 0$ & $\rightarrow$ & $\Lambda_{cd}=0$ &  &  \\ 
\hline
$\Lambda^{ab}$ & $G_{ab}=0$ & $\rightarrow$ &$\{G_{ab},L^-_{cd}\}\ne 0$  &  $\rightarrow$ &  $\lambda_{cd}=0$&  &   
\\
\hline
$u$ 
& ${\cal S}=0$ & $\rightarrow$ & $\Omega=0$ & $\rightarrow$  &$\{{\cal S},\Omega\}\ne 0$   &  $\rightarrow$ & u \\
\hline
\end{tabular} 
\caption{Variations over auxiliary variables and their dynamical consequences.}\label{table1}
\end{figure}

\section{Conclusion}
We demonstrated that tetrad approach allows to construct the canonical formalism for bigravity in the case of minimal potential   ($\beta_2=\beta_3=0$) by standard methods without complicated matrix transformations. We are sure that the same results can be obtained also for the general potential. The set of constraints and their algebra are analogous to those obtained in the metric approach, both  on the base of the Hassan-Rosen transformation~\cite{HR_bi}, and on the base of the axiomatic method~\cite{SolTch,Comelli_MayDay}. The key idea of the tetrad approach to bigravity was proposed by Rosen and Hinterbichler~\cite{HiRo}. In the present work we start with the tetrad formalism of GR developed in works by Deser, Isham, Nelson, Teitelboim, Henneaux and others~\cite{Deser_etal}. The similar approach were earlier proposed by Kluson~\cite{Kluson_tetrad}. Unlike Kluson's work we do not consider all variables appeared in the Lagrangian as canonical coordinates, but treat a great number of them (velocities of those are absent in the Lagrangian) as Lagrangian multipliers. It substantially reduces a number of constraints because a number of arising equations are treated as equations for Lagrangian multipliers, and not as constraints.   Though we believe that results should be independent on the subjective choice of variables, it seems that Kucha\u{r}'s method, i.e. exploiting two coordinate frames for space-time, the first one arbitrary and the second one determined by the family of hypersurfaces chosen, together with variables   $u=\bar N/N$ and $u^i=(\bar N^i-N^i)/N$ simplifies the problem considerably. The first significant difference between our approach and approach by Kluson is that we explicitely use the symmetry conditions (\ref{eq:condition}) and do not require the action to be stationary in varying over $p_a$. The second difference is that variable $u$ in this work is not free, but  fixed from constraint $\Omega$ consistency condition (\ref{eq:find_u}), whereas in article~\cite{Kluson_tetrad} variable $u$ is a Lagrangian multiplier standing before a first class constraint and so it is arbitrary. 

Another form of the GR action was taken as a starting point in works~\cite{Krasnov,Alexandrov} where the first order formalism taking connection as a canonical coordinate was used and on the base of this the canonical formalism of bigravity was constructed.  There is a similarity between the two approaches, but it is not so easy to demonstrate a one-to-one correspondence between them.   

We expect that the formalism derived here may be useful in studying fundamental issues of bigravity, such as existence of the partially massless case, causality problems, correspondence with the GR and so on.  

Finally we have added a Note on a recently proposed coupling to matter~\cite{dRHR}. There it is demonstrated that the Boulware-Deser ghost reappears for such coupling.

\section*{Appendix}
Expressions for $J^i$, $K^i$ are given by formulas:
\begin{eqnarray}
J^i
&=&  \frac{\partial{\cal S}}
{
\partial\bar{\cal H}_i
}
 =    \frac{\bar{\cal H}_k}{\beta_1e(1+\varepsilon)}
\left(
x^{(ik)}-\eta^{ik}
\frac{y}
{\varepsilon(1+\varepsilon)}\right)\nonumber
,\\
K^i&=&\frac{\partial{\cal R}'}{\partial\bar{\cal H}_i}=
\frac{\eta^{ij}\bar{\cal H}_j}{\beta_1e\varepsilon}.\label{eq:JK}
\end{eqnarray}
For $\Omega'$  we get
\begin{eqnarray}
\Omega'&=& K^j(P^{ma}e_{ja})_{,m}-P^{ma}K^je_{ma,j}\nonumber\\
&+&(J^mK^j\bar{\cal H}_j)_{,m}-J^m(K^j\bar{\cal H}_j)_{,m}+K^j(J^m\bar{\cal H}_m)_{,j}+(J^jK^m-K^jJ^m)\bar{\cal H}_{m,j}+\nonumber\\
&+&(Q^{ma}J^je_{ja})_{,m}+Q^{ma}J^je_{ma,j}-J^j(Q^{ma}e_{ja})_{,m}.
\label{eq:Omega'}
\end{eqnarray}
 where
\begin{eqnarray}
P^{ja}&=&\frac{\partial}{\partial{e^a_j}}({\cal S}-\bar{\cal H})=
e^{ja}\beta_1e
\left(
x-
\frac{y}
{\varepsilon(1+\varepsilon)} 
\right)
-e^{jb}\beta_1 e
\left(
x_{ba}+
\frac{y^{ba}}{1+\varepsilon} \right),\nonumber\\
Q^{ja}&=&\frac{\partial{\cal R}'}{\partial{e^a_j}}=
\frac{\beta_1ee^{ja}}{\varepsilon},\nonumber\\
T^{ja}&=&\frac{\partial{\cal S}}{\partial{f^a_j}}=
e^{ja}\beta_1e-f^{jb}\beta_1 e\frac{ y_{ab}}{1+\varepsilon}+\eta^{jk}\frac{\bar{\cal H}_k f^{ia}\bar{\cal H}_i y}{\beta_1e\varepsilon(1+\varepsilon)^2},\nonumber\\
Y^{ja}&=&\frac{\partial}{\partial{f^a_j}}({\cal R}'-{\cal H})=
-\frac{\eta^{jk}\bar{\cal H}_kf^i_a\bar{\cal H}_i}{\beta_1e\varepsilon}.\label{eq:varder}
\end{eqnarray}

\section*{Note added}
After completion of the above part of this work it was proposed~\cite{dRHR}  that all fields of matter in bigravity might interact with two gravitational fields  by means of the following combination of the two metrics (or two tetrads)
$$
G_{\mu\nu}=a^2g_{\mu\nu}+2abg_{\mu\alpha}\left(\sqrt{g^{-1}f}\right)^\alpha_{\ \nu}  +b^2f_{\mu\nu}\equiv (aE_\mu^A+bF_\mu^A)(aE_{\nu A}+bF_{\nu A}),
$$
but in this case Boulware-Deser ghost re-appeared. After a while in article~\cite{Hassan_etal} it was stated that really the number of degrees of freedom did not changed with a new interaction. 
This statement, in its turn, has been objected recently by the first group~\cite{dRHR2}. 
Let us consider
this problem in the approach derived above.


First,  mention that the new tensor $G_{\mu\nu}$ is symmetric due to symmetry conditions (7), and so may be called metric. 

Second,  decompose this new metric tensor over basis $(n^\mu,e^\mu_i)$ constructed with metric $f_{\mu\nu}$:
\begin{equation}
G_{\mu\nu}=G_{\perp\perp}n_\mu n_\nu +G_{\perp i}\left(n_\mu e_\nu^i+e_\mu^i n_\nu \right)+ G_{ij}e_\mu^i e_\nu^j,\label{eq:effmetric}
\end{equation}
where we have
\begin{eqnarray}
G_{\perp\perp}&\equiv& n^\mu n^\nu G_{\mu\nu}=a^2\left(-u^2+\gamma_{ij}u^i u^j \right)+2ab\left(-\varepsilon u +u^i e_{ia}p^a \right)-b^2,\nonumber\\
G_{\perp i}&\equiv& -n^\mu e^\nu_i G_{\mu\nu}=-a^2\gamma_{ij}u^j+2ab\left(uf_{ia}p_a-z_{ij}u^j\right),\nonumber\\
G_{ij}&\equiv&e^\mu_i e^\nu_j G_{\mu\nu}=a^2\gamma_{ij}+2abz_{ij}+b^2\eta_{ij}.
\end{eqnarray}
Let us introduce special notations $\psi_{ij}$ for the induced 3-metric $G_{ij}=\psi_{ij}$ and $\psi^{ij}$ for its inverse, so we have $\psi_{ik}\psi^{kj}=\delta_i^j$. 

Third, it is easy to provide the similar decomposition for inverse tensor $G^{\mu\nu}$: 
$$
G^{\mu\nu}=G^{\perp\perp}n^\mu n^\nu +G^{\perp i}\left(n^\mu e^\nu_i+e^\mu_i n^\nu \right)+ G^{ij}e^\mu_i e^\nu_j,
$$
where 
\begin{eqnarray}
G^{\perp\perp}&\equiv& n_\mu n_\nu G^{\mu\nu}=\frac{1}{G_{\perp\perp}},\nonumber\\
G^{\perp i}&\equiv& -n_\mu e_\nu^i G^{\mu\nu}=\frac{\psi^{ij}G_{\perp j}}{G_{\perp\perp}},\nonumber\\
G^{ij}&\equiv&e_\mu^i e_\nu^j G^{\mu\nu}=\psi^{ij}+\frac{G^{\perp i}G^{\perp j}}{G_{\perp\perp}}.
\end{eqnarray}

Then, interaction of matter with this metric can be illustrated by a simple example of the massless scalar field 
$$
{\cal L}^{(m)}=-\frac{1}{2}\sqrt{-G}G^{\mu\nu}\partial_\mu\phi\partial_\nu\phi.
$$
After Legendre transformation we arrive at the following canonical form
$$
{\cal L}^{(m)}=\pi\dot\phi-\hat N\hat{\cal H}^{(m)}-\hat N^i\hat{\cal H}_i^{(m)},
$$
where $$N^\mu\equiv\frac{\partial}{\partial t}=\hat N \hat n^\mu+\hat N^i e^\mu_i\equiv N n^\mu+ N^i e^\mu_i,$$ and 
$$
\hat{\cal H}^{(m)}=\frac{\pi^2}{2\sqrt\psi}+\frac{\sqrt\psi}{2}\psi^{ij}\partial_i\phi\partial_j\phi, \qquad \hat{\cal H}_i^{(m)}=\pi\partial_i\phi.
$$
Looking  at Eq.(10) we can write a similar formula
$$
\hat n^\mu=n^\mu\sqrt{-G^{\perp\perp}}
-e^\mu_i\frac{
G^{\perp i}
}{
\sqrt{-G^{\perp\perp}
}},
$$
and therefore
$$
\hat N=\frac{N}{
\sqrt{
-G^{\perp\perp}
}
}, \qquad \hat N^i=N^i-N\frac{
G^{\perp i}
}{
G^{
\perp\perp}
}.
$$
As a result we have
$$
{\cal L}^{(m)}=\pi\dot\phi-N\left(\sqrt{-G_{\perp\perp}}\hat{\cal H}^{(m)}-\psi^{ij}G_{\perp j}\hat{\cal H}_i^{(m)}\right)+ N^i\hat{\cal H}_i,
$$
or
$$
{\cal L}^{(m)}=\pi\dot\phi-N{\cal H}^{(m)}- N^i{\cal H}_i^{(m)},
$$
where
\begin{eqnarray}
{\cal H}^{(m)}&=&\frac{\sqrt{-G_{\perp\perp}}}{2}\left(\frac{\pi^2}{\sqrt{\psi}}+\sqrt{\psi}\psi^{ij}\partial_i\phi\partial_j\phi\right)+\nonumber\\
&+&\left(u^i-b^2u^k\eta_{kj}\psi^{ij}-2abu(f_jp)\psi^{ij}\right)\pi\partial_i\phi,\nonumber\\
{\cal H}_i^{(m)}&=&\hat{\cal H}^{(m)}_i.
\end{eqnarray}
Let us remind that according to Eq.(50) $\psi=\det|\psi_{ij}|$ and $\psi^{ij}$ depend on variables $f_{ia}$, $e_{ia}$ and $p_a$.

Now reconsider in brief the treatment given in Section 4. Hamiltonian (\ref{eq:H}) will now contain the explicit matter contribution 
$$
{\rm H}={\rm H}^{(f)}+{\rm H}^{(g)}+{\rm H}^{(m)}+\int d^3x \left(N\tilde U+\Lambda^aG_a+\Lambda^{ab}G_{ab}\right),
$$  
where
$$
{\rm H}^{(m)}=\int d^3x \left(N{\cal H}^{(m)}+N^i{\cal H}_i^{(m)}\right),
$$
Eqs. (18)--(23) will be here valid for gravitational contributions only.
Eq. (\ref{eq:H_0}) will take the following form 
\begin{eqnarray}
{\rm H}&=&\int d^3x \left(
N({\cal H}+u\bar{\cal H}+u^i\bar{\cal H}_i+\tilde U+{\cal H}^{(m)})+N^i({\cal H}_i+\bar{\cal H}_i+{\cal H}_i^{(m)})\right.+\nonumber\\
&+&\left.\lambda^+_{ab}L^+_{ab}+ \lambda^-_{ab}\bar L^-_{ab}+\Lambda^aG_a+\Lambda^{ab}G_{ab}\right).
\end{eqnarray}
Eqs. (25), (26) will be changed and become the following
\begin{eqnarray}
{\cal S}&\equiv&\frac{1}{N}\frac{\delta{\rm H}}{\delta u}= \bar{\cal H}+\frac{\partial(\tilde U+{\cal H}^{(m)})}{\partial u}=\nonumber\\
&=&\bar{\cal H}+\beta_1 e z +\frac{a^2 u+ab\varepsilon}{\sqrt{-G_{\perp\perp}}}\hat{\cal H}^{(m)}-\psi^{ij}f_{ja}p_a\hat{\cal H}_i^{(m)}
=0,\label{eq:Snew}\\
 {\cal S}_i&\equiv& \frac{1}{N}\frac{\delta{\rm H}}{\delta u^i}=\bar{\cal H}_i+\frac{\partial(\tilde U+{\cal H}^{(m)})}{\partial u^i}=\bar{\cal H}_i-\beta_1ef^a_ip_a-\nonumber\\
&-&\frac{a^2\gamma_{ij}u^j+abe_{ia}p_a}{\sqrt{-G_{\perp\perp}}}\hat{\cal H}^{(m)}+\left(a^2\gamma_{ij}+2abz_{ij} \right)\psi^{jk}\hat{\cal H}^{(m)}_i=0.\label{eq:Sinew}
\end{eqnarray}
We also should take into account symmetry conditions Eq.(\ref{eq:const1}), they does not change:
\begin{equation}
G_a\equiv p_a+up_bx_{ba}-u^jf_j^b\left(\delta_{ab}+\frac{p_ap_b}{\varepsilon+1}\right)=0.\label{eq:symm}
\end{equation}
To get a new constraint on the canonical variables  equations (\ref{eq:Snew}), (\ref{eq:Sinew}), (\ref{eq:symm})  have to be functionally dependent, i.e. Jacobian 
\begin{equation}
J=\frac{D({\cal S},{\cal S}_i,G_a)}{D(u,u^j,p_b)} =
\left(
\begin{array}{ccc}
 \frac{\partial{\cal S}}{\partial u}
& \frac{\partial{\cal S}}{\partial u^i} 
&\frac{\partial{\cal S}}{\partial p_a} \\
 \frac{\partial{\cal S}_j}{\partial u} 
& \frac{\partial{\cal S}_j}{\partial u^i}
& \frac{\partial{\cal S}_j}{\partial p_a}\\
 \frac{\partial{G_b}}{\partial u}
&\frac{\partial{G_b}}{\partial u^i}
& \frac{\partial{G_b}}{\partial p_a}
\end{array}
\right).\nonumber
\end{equation}
should be equal to zero. 

Derivatives of the above expressions over variables $u$, $u^i$ and $p_a$ are as follows:
\begin{eqnarray}
\frac{\partial{\cal S}}{\partial u}&=&
\frac{\partial^2(\tilde U+{\cal H}^{(m)})}{\partial u^2}=-\frac{a^2(u\gamma u)+2ab(uep)+b^2p^2}{(-G_{\perp\perp})^{3/2}} a^2\hat{\cal H}^{(m)},\nonumber\\
\frac{\partial{\cal S}}{\partial u^i}&=& 
\frac{\partial^2(\tilde U+{\cal H}^{(m)})}{\partial u\partial u^i}=\frac{\partial{\cal S}_i}{\partial u}=\frac{(au+b\varepsilon)(au_i+b(e_ip))}{(-G_{\perp\perp})^{3/2}} a^2\hat{\cal H}^{(m)},\nonumber\\
\frac{\partial{\cal S}}{\partial p_a}&=&
\frac{\partial^2(\tilde U+{\cal H}^{(m)})}{\partial u\partial p_a}=\frac{\beta_1 e}{\varepsilon+1}\left(x_{(ab)}p_b+\frac{yp_a}{\varepsilon(\varepsilon+1)} \right)+\nonumber\\
&+&ab\frac{\hat{\cal H}^{(m)}}{(-G_{\perp\perp})^{3/2}}\left((ue)_a(a^2u+ab\varepsilon)+\frac{p_a}{\varepsilon}(b^2-a^2(u\gamma u)-2ab(uep)+abu\varepsilon) \right)-\nonumber\\
&-&2abf_{ma}\psi^{im}\hat{\cal H}_i^{(m)}+\frac{2ab}{\varepsilon+1}\left(f_{ma}(e_np)+e_{na}(f_mp)-\frac{y_{mn}p_a}{\varepsilon(\varepsilon+1)} \right)\times\nonumber\\
&\times&\left(-\frac{\pi^2}{4\sqrt{\psi}}\psi^{mn}+(\psi^{mn}\psi^{ij}-\psi^{im}\psi^{jn})\frac{\sqrt{\psi}}{4}\partial_i\phi\partial_j\phi +2ab(f_jp)\psi^{im}\psi^{jn}\hat{\cal H}^{(m)}_i\right),\nonumber\\  
\frac{\partial{\cal S}_j}{\partial u^i}&=&\frac{\partial^2(\tilde U+{\cal H}^{(m)})}{\partial u^i\partial u^j}=
 -\frac{a^2\hat{\cal H}^{(m)}}{(-G_{\perp\perp})^{3/2}}\left(a^2u_iu_j+ab\left(u_i(e_jp)+u_j(e_ip)\right)+\right.\nonumber\\
&+&\left.b^2(e_ip)(e_jp)-\gamma_{ij}G_{\perp\perp}\right),
 \nonumber\\
\frac{\partial{\cal S}_j}{\partial p_a}&=&
\frac{\partial^2(\tilde U+{\cal H}^{(m)})}{\partial u^j\partial p_a}=-\beta_1 ef_{ja}+\nonumber\\
&+&ab\frac{\hat{\cal H}^{(m)}}{\sqrt{-G_{\perp\perp}}}\left(-e_{ja}+\frac{((ue)_a-u\frac{p_a}{\varepsilon})(a^2u_j+ab(e_jp))}{G_{\perp\perp}} \right)+\nonumber\\
&+&\frac{2ab}{\varepsilon+1}\left(f_{ma}(e_np)+e_{na}(f_mp)-\frac{p_a(f_mp)(e_np)}{\varepsilon(\varepsilon+1)} \right)\Biggl[b^2\eta_{ij}\psi^{im}\psi^{kn}\hat{\cal H}_k^{(m)}- \nonumber\\
&-&\frac{a^2\gamma_{ij}u^j+ab(e_ip)}{\sqrt{-G_{\perp\perp}}}\left(-\frac{\pi^2}{4\sqrt{\psi}}\psi^{mn}+(\psi^{mn}\psi^{ij}-\psi^{im}\psi^{jn})\frac{\sqrt{\psi}}{4}\partial_i\phi\partial_j\phi \right)\Biggr],\nonumber\\
\frac{\partial{G}_b}{\partial u}&=& 
p_ax_{ab},\nonumber\\
\frac{\partial{G}_b}{\partial u^i}&=&
f_i^a\left(\delta_{ba}+\frac{p_ap_b}{\varepsilon+1}\right) ,\nonumber\\
\frac{\partial{G}_b}{\partial p_a}&=&
\delta_{ab}\left(1-\frac{(ufp)}{\varepsilon +1}\right) +ux_{ab}-\frac{(uf)_ap_b}{\varepsilon+1}+\frac{(ufp)p_a p_b}{\varepsilon(\varepsilon +1)^2}.
\end{eqnarray}
Here we use some new notations: $(u\gamma u)=\gamma_{ij}u^iu^j$, $(uep)=u^ie_{ia}p_a$, $(e_ip)=e_{ia}p_a$, $(f_ip)=f_{ia}p_a$, $(ue)_a=u^ie_{ia}$, $(uf)_a=u^if_{ia}$.

One can see that Jacobian (58) is a polynomial in $\hat{\cal H}^{(m)}$ and $\hat{\cal H}_i^{(m)}$.  Then it is impossible for this Jacobian to be identically zero for arbitrary values of the scalar field. Therefore Eqs.(55)-(57) are here not constraints but equations to be solved for auxiliary variables $u$, $u^i$ and $p_a$. We are to acknowledge that the Boulware-Deser ghost can not be avoided for the coupling of matter to effective metric (\ref{eq:effmetric}).

\end{document}